\documentclass[conference]{IEEEtran}
\usepackage{verbatim} 
\usepackage{graphicx}
\usepackage{subfigure}
\usepackage{times}
\usepackage{caption}
\usepackage{epsfig}
\usepackage{stfloats}
\usepackage[noadjust]{cite}
\usepackage{enumerate}
\usepackage{amssymb}
\usepackage{multirow}
\usepackage{latexsym}
\usepackage{color}
\usepackage{amsmath}
\usepackage{lipsum}
\usepackage[noend]{algpseudocode}
\usepackage{epstopdf}
\usepackage{tikz}
\usepackage[american]{circuitikz}
\usepackage{algorithm}
\usepackage{mathtools}

\allowdisplaybreaks
\usepackage{etoolbox}

 \makeatletter 
 \def\@eqnnum{{\normalsize \normalcolor (\theequation)}} 
  \makeatother
\usepackage{etex}
\usepackage{tabularx} 
\usepackage{algorithm}
\usepackage{graphicx}
\usepackage{algpseudocode}
\usepackage{tikz}
\usetikzlibrary{shapes,arrows}
\usepackage{pstricks}
\usepackage{pst-node,pst-blur}
\usetikzlibrary{arrows}
\usepackage{amsfonts}
\usepackage{bm}
\usepackage{tabularx}
\usepackage{enumerate}
\usepackage{subfigure}
\usepackage{amsmath}
\usepackage{amssymb}
\usepackage{booktabs} 
\usepackage{multirow} 
\usepackage{mathtools} 
\usepackage{epsfig}
\usepackage{epstopdf}
\usepackage{array}

\usepackage[noadjust]{cite}
\usepackage{caption}
\usepackage{etoolbox}

\allowdisplaybreaks
\makeatletter
\def\set@curr@file#1{%
  \begingroup
    \escapechar\m@ne
    \xdef\@curr@file{\expandafter\string\csname #1\endcsname}%
  \endgroup
}
\def\quote@name#1{"\quote@@name#1\@gobble""}
\def\quote@@name#1"{#1\quote@@name}
\def\unquote@name#1{\quote@@name#1\@gobble"}
\makeatother
\usepackage{graphics}

\begin{document}

\title{Power-efficient Joint Link Selection and Multi-hop Routing for Throughput Maximization in\\ UAV Assisted FANETs}
\author{\IEEEauthorblockN{Payal Mittal, Santosh Shah and Anirudh Agarwal }
\IEEEauthorblockA{Department of Electronics and Communication Engineering\\ 
The LNM Institute of Information Technology, Jaipur, Rajasthan, India\\
%\IEEEauthorrefmark{2}School of Electrical Engineering and Telecommunications, University of New South Wales, Sydney, Australia
\\E-mails: \{payalmittal.y19, santosh.shah, anirudh.agarwal\}@lnmiit.ac.in
}}

\maketitle

\begin{abstract}
This paper considers a multi-UAV network with a ground station (GS) that uses multi-hop relaying structure for data transmission in a power-efficient manner. The objective is to investigate the best possible multi-hop routing structure for data transmission to maximize the overall network throughput of a flying ad-hoc network (FANET) of UAVs. We formulate a problem to jointly optimize the multi-hop routing structure with the communication link selection for a given power budget so that the overall network throughput can be maximized. It appears that the formulated problem belongs to a class of nonconvex and integer optimization problems, thus making it NP-hard. To solve this problem efficiently, it is decoupled into two subproblems $\textbf{i)}$ power allocation with known Bellman Ford-based multi-hop routing structure and $\textbf{ii)}$ link selection problem. Further, these two subproblems are independently converted into convex problems by relaxation and solved in tandem for the best suboptimal solution to the main problem. Simulation results indicate that the proposed multi-hop routing schemes can achieve a significant improvement in network throughput compared to the other benchmark scheme.

% \textcolor{blue}{Simulation results verify a significant increase in throughput of FANET by exchanging appropriate child-parent links of UAV in the link-selection sub-problem.}

\end{abstract}
 %\vspace{-1.5mm}
% \IEEEpeerreviewmaketitle
% 
% \vspace{-3mm}
% \section{Introduction}
\section{Introduction}
In recent years, unmanned aerial vehicles (UAVs), also known as a drones, are gaining popularity for their significant potential uses in wireless communication due to their high mobility, high flexibility, and high adaptability. The UAV is a rapidly growing market that has already found many applications in military, civilian and public domains \cite{mozaffari2019tutorial}. Recently, the network of multiple UAVs grouped in an ad-hoc manner has attracted significant attention for multi-hop communication to extend the coverage during an emergency situation. However, UAVs are commonly deployed for aerial communication, and monitoring in some crisis circumstances, such as earthquakes and floods \cite{mayor2019deploying, Merwaday}. The network in which several UAVs can share data and collaborate in an ad-hoc manner is a flying ad-hoc network (FANET) in which each UAV can also operate as a relay \cite{UAV-Relaying}. Generally, FANETs comprise of a ground station (GS) and UAVs hovering and flying at a particular permissible altitude. A possible FANET scenario where all UAVs communicate bidirectional to share their information and also act as relays during information transfer to the GS via multi-hop. Due to the highly mobile and dynamic nature of UAV ad-hoc networks, a routing protocol is required to tackle the collision and interference issues. Cooperative relaying using UAVs with an efficient multi-hop routing structure provides more coverage, reliable data transmission, enhanced data rates, and better network connectivity. However, few challenges remain in FANETs such as throughput maximization and power-efficient multi-hop routing with optimal link selection, which are required to be jointly optimized.

% \textcolor{blue}{The unmanned aerial vehicle (UAV), also known as a drone, is a rapidly growing market that has already found many applications in military and civilian domains such as emergency communications, disaster relief, surveillance, reconnaissance operations, hazardous site inspection, search and rescue operations \cite{mozaffari2019tutorial}. However, UAVs are commonly deployed for aerial communication, and monitoring in some crisis circumstances, such as earthquakes and floods \cite{mayor2019deploying, Merwaday}.} 

\subsection{State-of-the-Art}
Various routing techniques have been introduced in recent years for effective data collection and dissemination. The concept of UAV-assisted cooperative communication with a multi-hop routing structure has been well demonstrated in the existing literature \cite{rosatirouting, lakew_routing, sang_routing, Zhang_routing}. In \cite{rosatirouting}, authors proposed a predictive optimized link-state routing protocol, which takes the advantages of global positioning system information to predict the quality of wireless channel to find the routing with minimum interruptions and delays. To deal with UAV's limited energy resources and storage capacity, the authors in \cite{sang_routing} proposed a multi-hop routing technique based on trajectory prediction. Further in \cite{Zhang_routing}, the author introduced a packet arrival prediction routing protocol to improve the link reliability. Moreover, an iterative distributed algorithm is proposed in \cite{shah2013joint} for multi-hop routing, which provides a trade-off between energy efficiency and estimation accuracy.

Furthermore, several important works investigate the problem of throughput maximization and power allocation. Specifically, \cite{Ono_rate} proposed a variable rate relaying approach for fixed-wing UAVs to optimize the achievable rate of the system. Further, the authors in \cite{zeng2016throughput} proposed a novel framework to maximize the system throughput by jointly optimizing the power allocation and trajectory of a single UAV-based mobile relaying network. Similarly, \cite{fan2018optimal} jointly optimized the bandwidth, transmission power, transmission rate, and  UAV’s position for maximizing the system throughput. Moreover, the authors in \cite{agarwal2020altitude} optimized the UAV's altitude by considering the problem of minimizing the network outage probability. Recently, the authors in \cite{Gupta_trajectory} focused on trajectory optimization for a UAV-assisted communication to maximize the average sum rate of all the users.

\subsection{Research Gap and Motivation}
Most of the above works \cite{rosatirouting, sang_routing, Zhang_routing} deal with different routing protocols to optimize the resource allocation in different ways but the discussion on multi-hop routing structure with efficient power allocation to maximize the network throughput has not been covered yet in the current literature. Further, the authors in \cite{Ono_rate, zeng2016throughput} consider a single fixed-wing UAV for optimizing the system throughput. However, a single fixed-wing UAV in the event of a disaster (such as a flood, earthquake, or other natural disasters) may not be a productive choice where continuous monitoring is needed. In this case, rotary-wing UAVs provide a number of advantages over a fixed-wing UAV, including better maneuverability, payload capacity, and cost-effective design. Furthermore, the authors in \cite{fan2018optimal, agarwal2020altitude} considered only one rotary-wing UAV to serve the single and multiple communication pairs on the ground, respectively, while neglecting the multi-hop communication among the UAVs. Similarly, \cite{Chen_Multihop} considered the multi-hop single link and multiple dual-hop links between transmitter and receiver but did not consider the multi-hop multi-link to determine the system's performance.

From the above discussion, it can be observed that it is critical to design efficient routing structure with proper network flow for an efficient corporation and information exchange among multiple UAVs over multiple hops in a FANET. System performance can be further improved by appropriate power allocation. Therefore, a multi-hop routing structure with reliable and efficient power allocation to maximize the overall network throughput needs further studies. So in this work, we aim to maximize the network throughput of a FANET by jointly optimizing the UAV’s power allocation along with the multi-hop routing structure.

\subsection{Novelty and Scope}
{To the best of our knowledge, this is unexplored work that considers throughput maximization while optimizing a multi-hop routing structure for a given total power budget in multi UAV-assisted FANETs}. Disaster management, rescue agencies, public safety bodies, and defense organizations may get benefited from the proposed framework for practical applications.

\subsection{Major contributions}
The key contributions of this work are: 
\begin{itemize}
\item The unique network model is proposed in Section II, in which we formulate an optimization problem to maximize network's overall throughput while considering power allocation and communication link selection as two different variables. 
\item We have shown that this problem is nonconvex integer optimization problem, thus making it NP-hard.
\item  In order to solve this nonconvex integer optimization problem, it is decoupled into two subproblems. The first subproblem optimizes power allocation while using Bellman Ford based routing algorithm and second subproblem optimizes the multi-hop routing structure while using the solution of the first subproblem, which are described in Section III.
\item These two subproblems are solved in tandem to find the best global sub-optimal closed-form solution to the original problem. Then, this closed-form solution is used to construct the best routing among UAVs and from UAVs to the GS to obtain the best possible overall throughput.
\item  In Section IV, we provide the simulation results, followed by the conclusion in Section V.
\end{itemize}
%\emph{Notations:} The following notations are used throughout the paper. Bold uppercase letters denote matrices and bold lowercase letters denote the vectors. $\left [ \mathbf{X}_{ij} \right ]$ denotes the $(i,j)^{th}$ entry of matrix $\mathbf{X}$. 

\section{Network Model and Problem Formulation}

\subsection{Network Model}
\begin{figure}[t]
\centering
\includegraphics[width=3.48in, height=2.5in]{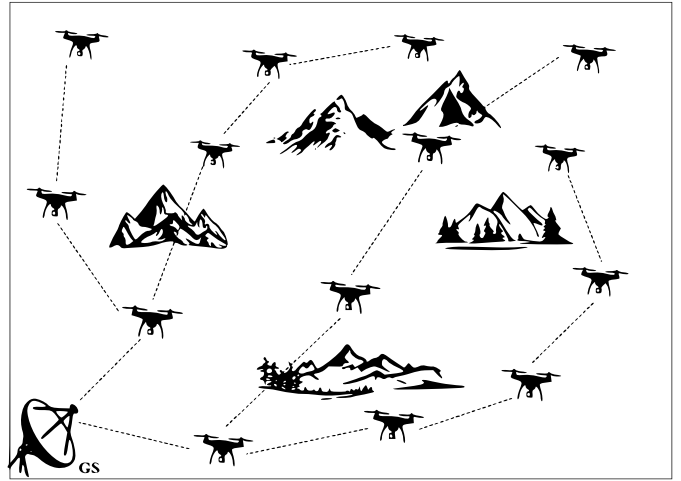}
\caption{Illustration of a UAV-assisted FANET with multi-hop routing structure considered in this work. All the monitored information reach to the ground station (GS) using the multi-hop routing, where intermediate UAVs act as a relay.}
\label{fig1:system model}
\end{figure}
We consider a scenario of UAV-assisted FANET, where multiple UAVs communicate in a multi-hop manner as shown in Fig \ref{fig1:system model}. This system model can be used in a high terrain environment where a disaster (such as an earthquake, flood, or other natural disasters) may occur. We assume that the FANET consists $n$ rotary-wing UAVs, $\mathbf{u} \in$ [$u_{1}, u_{2}, \cdots u_{n} $] that are randomly deployed to monitor over the disaster-prone square area of approximately $20\times 20$ km$^{2}$. We consider that each UAV is equipped with a camera, an image encoder, and a radio transceiver. We consider one ground station (GS) to receive all information picked up by UAVs. Long-distance communication between far-away UAVs and GS is not possible to establish a power-efficient FANET; however, UAVs that are in proximity to GS may communicate directly with GS. UAVs can hover and stay stationary over a given area for a particular amount of time. The sensed data is communicated to GS using UAVs via multi-hop relaying.

We assume that sufficiently charged UAVs will be on standby to replace the low-power UAVs in real-time. Furthermore, we have assumed that the distance between UAVs is large enough to prevent collisions and interference \cite{khuwaja2019optimum}. A Cartesian coordinate system is considered for simplicity of analysis. The coordinates of the $n$-th UAV are $(x_{n},y_{n},z_{n})$. A UAV that captures the information becomes the source node and the information is transmitted to GS directly or via multi-hop, whichever is application. The coordinates of GS is given by $(x_{n+1}, y_{n+1},0)$. The altitude of a UAV remains constant once it has reached its fixed position. We assume all the UAVs fly at a fixed permissible altitude to provide a line-of-sight connection.

$L_{ij}$ is the communication link that exists between $i^{th}$ and $j^{th}$ UAVs. We assume all UAVs are at the same height, then the distance $d_{ij}$ is given as,  
\begin{equation}\label{dij}
d_{ij} = \sqrt{(x_{i}-x_{j})^{2}+(y_{i}-y_{j})^{2}},
\end{equation}
\noindent where $i \in \{1,2,..., n\} \triangleq \mathbb{N}_0 \hspace{2mm} \text{and}\hspace{2mm} j \in \{1,2,..., n+1\} \triangleq \mathbb{N}$, $j$ is an expected parent of $i$ and $n+1$ indicates the coordinate of GS. Furthermore, the underlying channel gain $h_{ij}$ is defined as,
\begin{equation}\label{Hij}
h_{ij} = \frac{\alpha_{0}}{\left ({d_{ij} }\right)^{\beta}},
\end{equation}
\noindent where $\alpha_{0}$ is the received power at a reference distance $d_{0}=1 \;\textrm{m}$, $\beta$ is the free-space path loss exponent.

We use $P_{ij}$ to denote the transmission power for the link $L_{ij}$, $B$ is the channel bandwidth and $\sigma^{2}$ is the noise spectral density, then capacity of the link $L_{ij}$ can be expressed as,
\begin{equation}\label{Cij}
C_{ij}  = B \log_2 \left( 1+\frac{P_{ij}h_{ij}}{\sigma^{2} B} \right).
\end{equation}
Further, the transmission rate of the link $L_{ij}$ is $R_{ij}$, then 
\begin{equation}\label{Rij}
R_{ij}\leq B \log_2\left(1+\frac{P_{ij}h_{ij}}{\sigma^{2} B}\right),\,\hspace{4pt} \forall i \, \in \mathbb{N}_0,\hspace{1mm} \forall j \in \mathbb{N}.
\end{equation}

\subsection{Optimization Problem Formulation}
In this work, we aim to maximize the overall network throughput $\sum_{i=1}^{n} R_{ij}$ by optimizing both variables $P_{ij}$ and $L_{ij}$. The data collected by UAVs should be routed to the GS via power-efficient multi-hop routing paths. The optimization problem \text{($\bm{\mathbb{P}}1$)} can be formulated as follows,\\ 
$
\begin{aligned}
\hspace{2mm} \text{($\bm{\mathbb{P}}1$):}  \hspace{2mm}
\hspace{0mm} &\underset{P_{ij},\hspace{1mm} L_{ij}}{\text{maximize}} \sum_{i=1}^{n}\sum_{j=1}^{n+1}L_{ij}R_{ij}\\
& \textrm{subject to} \hspace{2mm}C1\hspace{-1mm}: \sum_{i=1}^{n}\sum_{j=1}^{n+1}L_{ij} P_{ij} =  P_{b}, 
\hspace{4mm} 
\\ 
& \hspace{17mm} C2: P_{ij} \ge 0, \, \hspace{4pt} \forall i \in \mathbb{N}_0,\hspace{1mm} \forall j \in \mathbb{N},\\
&\hspace{17mm} C3: L_{ij} \in\{0,1\}, \, \hspace{4pt} \forall i \in \mathbb{N}_0,\hspace{1mm} \forall j \in \mathbb{N}, \hspace{12mm} \\
& \hspace{17mm} C4\hspace{0mm}: {L_{ij}}\leq {L_{jk}},  \hspace{4pt} \forall i  \in \mathbb{N}_0,\hspace{1mm} \forall j, k \in \mathbb{N}
,\\
&\hspace{17mm} C5:  {L_{ij}}+{L_{ji}}\leq 1,  \: \hspace{4pt} \forall i \neq j , \hspace{12mm} \\
& \hspace{17mm} C6\hspace{0mm}: A_{ij}= 1, \,A_{ij}\in\mathbf{A},\forall i \in \mathbb{N}_0,\hspace{1mm} \forall j \in \mathbb{N}, \hspace{4pt} 
\end{aligned}
$

% \noindent where $\mathbb{N}\in \mathbb{N}_0 \cup \{n+1\}$. 

\par The sum of associated maximum power for each UAV should be restricted to have some communication power budget $P_{b}$ and the power associated with each UAV cannot be negative according to the constraints $C1$ and $C2$, respectively. The constraint $C3$ represents that $L_{ij}$ is an integer variable indicating the status of selecting a particular link between  $i^{th}$ and $j^{th}$ UAV, i.e., $L_{ij} = 1$ denotes that the link is selected for communication and $L_{ij} = 0$ denotes that the link is inactive. The constraint $C4$ ensures that the direction of data flow is towards GS, where parent $j$ of $i$ and parent $k$ of $j $ are chosen such that the resulting information is rooted at GS. Finally, $C5$ prevents the loops in a multi-hop routing path and allows the routing to head in the correct direction. $\mathbf{A}$ denotes the incidence matrix with elements $A_{ij}$, represented by constraint $C6$.
The elements of incidence matrix $\mathbf{A}$ are given as
\begin{equation}
A_{ij} = \begin{cases}
1 & \text{if} \; \; \;d_{ij}\leq d_{th} \\ 
0 & \text{ otherwise,} \\ 
\end{cases}
\end{equation}
where $d_{th}$ is a maximum allowed threshold distance for any two UAVs to communicate to avoid bad channel communications. The problem \text{($\bm{\mathbb{P}}1$)} is a nonconvex integer optimization problem \cite{boyd2004convex}, thus it is hard to solve in its original form. To solve this problem, one variable is considered at a time resulting in two subproblems, which are then solved efficiently in tandem to get the near-optimal solution of the original problem \text{($\bm{\mathbb{P}}1$)}.

 \section{Proposed Optimal Solution Methodology}
In the following, we consider two subproblems of \text{($\bm{\mathbb{P}}1$)}, namely power allocation subproblem and link selection subproblem to optimize multi-hop routing.

\subsection{Power Allocation Subproblem}
\par In this section, we consider the first subproblem of \text{($\bm{\mathbb{P}}1$)} to optimize power allocation for each UAV by assuming that a multi-hop routing structure based on Bellman Ford shortest path tree (SPT) algorithm is given. As a result, this subproblem is only with the optimization variable $P_{ij}$, whereas $L_{ij}$ is known by using SPT, in the sense that a parent of $i$, i.e., $j$ is known. Hence, index $j$ from \text{($\bm{\mathbb{P}}1$)} is omitted for brevity. So, the power allocation subproblem  can be reformulated as,
\\ 
$\begin{aligned}
\hspace{11mm} \text{($\bm{\mathbb{P}}1.1$)}: \hspace{2mm} &\underset{P_{i},\hspace{1mm} SPT}{\text{maximize}} \sum_{i=1}^{n}R_{i} \\
&  \textrm{subject to}\hspace{2mm}C1: \sum_{i=1}^{n} P_{i} =  P_{b},  \hspace{6mm} \\& \hspace{16mm}C2: P_{i} \ge 0, \, \hspace{4pt} \forall\hspace{2pt} i \in \mathbb{N}_0.
\end{aligned}
$\\\\
Thus, there exists an optimal solution to \text{($\bm{\mathbb{P}}1.1$)} such that both the constraint $C1$ and $C2$ satisfied. Note that \text{($\bm{\mathbb{P}}1.1$)} is a convex optimization problem, which can be solved numerically using standard convex optimization techniques such as Lagrangian method and generates the optimal solution as $\{P_{i}^{*}\}$.\\
\emph{Proof}: \textit{Refer to Appendix for convexity.}
\\
\par The Lagrangian $\mathcal{L}$ of \text{($\bm{\mathbb{P}}1.1$)} is expressed as, 

\begin{equation}\label{lagrange}
    \begin{split}
 \mathcal{L}(P_{i}, \lambda, \mu_i) = R - \lambda \left(\sum_{i=1}^{n} P_{i} - P_{b} \right) + \sum_{i=1}^{n} \mu_i P_{i},
    \end{split}
    \end{equation}
where $R \triangleq \sum_{i=1}^{n} R_i$, $\lambda$  and $\mu_i\, \forall i \in \mathbb{N}_0$ are the non-negative Lagrange multipliers with respect to $C1$ and $C2$ respectively. As \text{($\bm{\mathbb{P}}1.1$)} is convex, the global optimal solution is provided by the Karush-Kuhn-Tucker (KKT) point $(P_{i}^{*}, \lambda^{*}, \mu_i^{*})$. The KKT conditions are as follows:

\begin{subequations}
\begin{align}
&\frac{\partial\mathcal{L}}{\partial P_{i}} \triangleq \frac{Bh_{i}}{\sigma^{2}B+P_{i}h_{i}}-\lambda+\mu_i = 0,   \hspace{4pt} \forall i \in \mathbb{N}_0,\label{eq:KKT1}\\
&\frac{\partial\mathcal{L}}{\partial \lambda } \triangleq \left(\sum_{i=1}^{n} P_{i} - P_{b} \right) = 0, \hspace{4pt} \forall i \in \mathbb{N}_0,\label{eq:KKT2}\\
&\frac{\partial\mathcal{L}}{\partial \mu_i} \triangleq \left(\sum_{i=1}^{n} P_{i} \right) = 0, \hspace{4pt} \forall i \in \mathbb{N}_0.\label{eq:KKT3}
\end{align}
\end{subequations}
\par Without loss of generality, we take $\mu_i = 0$ because $P_{i}>0$  $\forall \,i\in \mathbb{N}_0$. Then, after solving (\ref{eq:KKT1}) we get,

\begin{equation}\label{Pi1}
 P_{i}^{*} = \frac{B}{\lambda}  - \frac{\sigma ^{2}B}{h_{i}}.
\end{equation}

As $P_{i}>0$, $\frac{B}{\lambda} - \frac{\sigma ^{2}B}{h_{i}}$ should be positive $\forall i\in \mathbb{N}_0$. Because subtraction of two quantities can only provide power if each individual quantity is power, the terms $\frac{B}{\lambda}$ and $\frac{\sigma ^{2}B}{h_{i}}$ can be considered as powers. So in nutshell, 

\begin{equation}\label{Pi2}
P_{i}^{*} = \left\{\begin{matrix}
\frac{B}{\lambda}  - \frac{\sigma ^{2}B}{h_{i}}, &    \; \lambda < \frac{h_{i}}{\sigma ^{2}}\\ 
0, &  \; \lambda \geq \frac{h_{i}}{\sigma ^{2}}
\end{matrix}\right.
\end{equation}
$\mu_{i}^{*} = 0 $
% \begin{equation}\label{Pi3}
% P_{i}^{*} = \text{max}\left [ 0, \;  \frac{W}{\lambda}  - \frac{\sigma ^{2}W}{h_{i}}\right ]. 
% \end{equation} 
Finally, by solving (\ref{eq:KKT2}) and (\ref{Pi2}), we get
\begin{equation}\label{lambda}
\lambda^{*} = \frac{n}{\frac{P_{b}}{B}+\sum_{i=1}^{n}\frac{\sigma^{2}}{h_{i}}}.
\end{equation}
\subsection{Link Selection Subproblem}
In this section, we consider second subproblem of \text{($\bm{\mathbb{P}}1$)} to optimize link selection $L_{ij}$, that is this subproblem further optimizes the SPT based routing structure to enhance the overall network throughput given the solution of \text{($\bm{\mathbb{P}}1.1$)}. The optimal objective value $\{P_{i}^{*}\}$ of the optimization problem \text{($\bm{\mathbb{P}}1.1$)} will be used as the solution base for this problem. Then, the link selection subproblem can be written as $\text{($\bm{\mathbb{P}}1.2$)}$,
$
\begin{aligned}
\hspace{2mm} \text{($\bm{\mathbb{P}}1.2$):}  \hspace{2mm}
\hspace{0mm} &\underset{\{L_{ik}\}}{\text{maximize}} \sum_{i=1}^{n}\sum_{k\in\mathcal{N}_i, k\neq j}L_{ik}R_{ik}\\
& \textrm{subject to} \hspace{2mm}C1\hspace{-1mm}: \sum_{i=1}^{n}\sum_{k\in\mathcal{N}_i, k\neq j} L_{ik} {P_{ik}} =  P_{i}^{*}, 
\hspace{4mm} 
\\
&\hspace{16mm} C2: L_{ik} \in\{0,1\}, \, \hspace{4pt} \forall i  \, \in \mathbb{N}_0,\hspace{1mm} \forall k \in \mathbb{N}_i, \hspace{12mm} \\&\hspace{16mm} C3\hspace{0mm}: {L_{ik}}\leq {L_{ki}}, \hspace{4pt} \forall i \, \in \mathbb{N}_0,\hspace{1mm} \forall k \in \mathbb{N}_i
,\\
&\hspace{16mm} C4: {L_{ik}}+{L_{ki}}\leq 1, \hspace{4pt} \forall i \neq k , \hspace{12mm} 
\end{aligned}
$\\
where index $k$ indicates all possible one-hop neighbours $\mathcal{N}_i$ of UAV $i$ that can be chosen as the next best parent of $i$, if for the same $P_i^*$, $R_{ik}>R_{ij}$ ($j$ is the previous parent of $i$). 

% \begin{equation}
% P_{ij}^{'} = \begin{bmatrix}
%  \overline{P_{i}^{*}}& \overline{P_{i}^{*}} & \cdots  & \overline{P_{i}^{*}}
% \end{bmatrix} 
% \end{equation}

\par \text{($\bm{\mathbb{P}}1.2$)} is a non-convex optimization problem due to the non-convex constraint $C2$, making it an NP-hard integer optimization problem. Therefore, the optimal solution is not feasible. This problem is solved by executing an approximate relaxation over the variable and then solving the relaxed problem. We can cast this problem into an equivalent problem formulation by relaxing $C2$ constraint so that it can take values anywhere in the range $0 \leq L^r_{ik}\leq 1$. 

Let $D_{ik} \in \mathbf{D}$ be the elements of network's directional matrix. so that 

\begin{equation}
D_{ik} = \begin{cases}
1 & \text{ if $k$ is the parent of $i$ }  \\ 
-1 & \text{ if $i$ is the parent of $k$} \\ 
0 & \text{ otherwise } \\ 
a>1 & \text{ if there is link exists between $i$ and $k$, } 
\end{cases}
\end{equation}
where $a$ is positive rational integer. With this notation, we can reformulate problem \text{($\bm{\mathbb{P}}1.2$)} as follows:
\\
$
\begin{aligned}
\hspace{2mm} \text{($\bm{\mathbb{P}}1.3$):}  \hspace{2mm}
\hspace{0mm} &\underset{\{L_{ik}^{r}\}}{\text{maximize}} \sum_{i=1}^{n}\sum_{k\in\mathcal{N}_i, k\neq j}L_{ik}^{r}R_{ik}D_{ik}\\
& \textrm{subject to} \hspace{2mm} C1\hspace{-1mm}: \sum_{i=1}^{n}\sum_{k\in\mathcal{N}_i, k\neq j}L_{ik}^{r} P_{ik} =  P_{i}^{*}, 
\hspace{4mm} 
\\
&\hspace{17mm} C2: 0\leq L_{ik}^{r}\leq 1 \,\hspace{4pt} \forall i\in \mathbb{N}_0,\hspace{2pt} \forall k\in \mathbb{N}_i,\hspace{12mm}
\end{aligned}
$\\ \\
where $L_{ik}^{r}\in [0,1]$ is the relaxed version of variable $L_{ik}\in \{0,1\}$. In this case, the objective function is convex function over $L_{ik}^{r}$ and all other constraints are linear over $L_{ik}^{r}$, resulting in a well-defined convex problem. Now, we describe a simple method to solve the relaxed problem \text{($\bm{\mathbb{P}}1.3$)} very efficiently but approximately using \textit{log barrier method} \cite{boyd2004convex}, that is, the interior point method is used to solve this problem. Then, the problem \text{($\bm{\mathbb{P}}1.3$)} can also be posed as:
\\
$
\begin{aligned}
\hspace{2mm} \text{($\bm{\mathbb{P}}1.4$):}  \hspace{2mm}
\hspace{0mm} &\underset{\{L_{ij}^{r}\}}{\text{maximize}} \hspace{2mm} \phi \left ( \mathbf{L^{r}} \right )\\
& \textrm{subject to} \hspace{2mm} C1\hspace{-1mm}: \sum_{i=1}^{n}\sum_{k\in\mathcal{N}_i, k\neq j}L_{ik}^{r} P_{ik} =  P_{i}^{*}. 
\hspace{4mm} 
\end{aligned}
$\\ \\
where  $\phi \left (\mathbf{L^{r}} \right ) = \sum_{i=1}^{n}\sum_{k\in\mathcal{N}_i, k\neq j}L_{ik}^{r}R_{ik}D_{ik} + \frac{1}{\gamma} \sum_{i=1}^{n}[\log(L_{ik}^{r})+\log(1-L_{ik}^{r})]$ and  $\gamma > 0$ to set the quality of approximations. The optimization problem \text{($\bm{\mathbb{P}}1.4$)} is with concave objective and equality constraint is linear, therefore it can be efficiently solved by the Newton's method \cite{boyd2004convex}. In this method, at each step Newton search step $\Delta \mathbf{L^r}$ is computed, which is expressed by
\begin{equation}
\Delta \mathbf{L^{r}} = \left (\bigtriangledown^{2}\phi  \right )^{-1}\bigtriangledown\phi - \Bigg(\frac{\mathbf{P_i}^T\left (\bigtriangledown^{2}\phi  \right )^{-1}\bigtriangledown\phi}{\mathbf{P_i}^T\left (\bigtriangledown^{2}\phi  \right )^{-1} \mathbf{P_i}} \Bigg)\left (\bigtriangledown^{2}\phi  \right )^{-1} \mathbf{P_i} 
\end{equation}
where $\bigtriangledown\phi$ and $\bigtriangledown^{2}\phi$ are the gradient and Hessian of function $\phi$, respectively. We take $\text{diag} (\mathbf{L^r})\mathbf{P_i}=P_i^*\mathbf{1}$ as initial point, where $\mathbf{P_i}^T = [P_{i1}, P_{i2}, \dots, P_{ik}]$; a column vector with elements as all possible parent's link power.

The backtracking line search is then used to take the equality constraint into account and update $\mathbf{L^{r}}$ by replacing it with $\mathbf{L^{r}}+\tau \mathbf{\Delta L^{r}}$, where $\tau \: \epsilon \: \left ( 0,1   \right ]$ is step size for backtracking line search. We stop when the Newton decrement $\left (-\bigtriangledown \phi \left ( \mathbf{L^{r}} \right )^{T} \Delta \mathbf{L^{r}}  \right )^{1/2} \leq \varepsilon$, for $\varepsilon > 0$ sufficiently small. 
In our problem, for completeness we provide expressions for the first and second derivatives of $\phi$ in terms of its gradient $\bigtriangledown \phi $ and the Hessian $ \bigtriangledown^{2} \phi $, which can be written as: \\
\begin{equation}\label{gradient}
(\bigtriangledown\phi)_i =  \sum_{k\in\mathcal{N}_i, k\neq j}R_{ik}D_{ik}+\frac{1}{\gamma}\Bigg(\frac{1}{L_{ik}^{r}} - \frac{1}{1-L_{ik}^{r}}\Bigg). 
\end{equation} 
The Hessian $ \bigtriangledown^{2} \phi $ can be written as: 
\begin{equation}\label{Hessian}
(\bigtriangledown^{2}\phi)_i = -\frac{1}{\gamma}\Bigg(\frac{1}{(L_{ik}^{r})^{2}} + \frac{1}{(1-L_{ik}^{r})^{2}}\Bigg).
\end{equation}

The solution of \text{($\bm{\mathbb{P}}1.4$)} that is $\{L_{ik}^{r*}\}$ generated from the above procedure and the solution $\{P_i^*\}$ of \text{($\bm{\mathbb{P}}1.1$)} are then used to update the overall network throughput.

\section{Results and Performance Evaluation}
This section deals with the validation and other numerical results along with the key optimal insights. Unless explicitly stated, we have considered, $B=10$ MHz as the total system bandwidth, $\alpha _{0}$ is set to be $\left(\frac{c}{4\pi f}\right)^{2}$, where $c$ is the speed of light and the centre frequency $f=1$ GHz. The noise power spectral density corresponds to $\sigma^{2}=-174$ dBm/Hz. The UAVs are placed such that the minimum altitude $H$ is $150$ m. All the results are generated by averaging over 100 iterations with different random seeds.

\begin{figure}[h]
\centering
\includegraphics[width=3.8in]{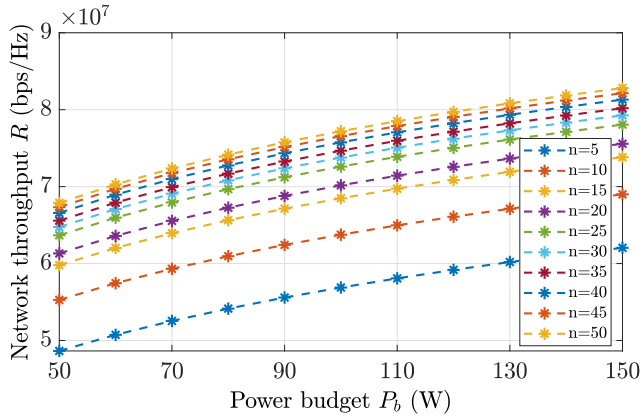}
\caption{Impact of power budget $P_{b}$ on network throughput $R$ for different number of UAVs ($n$) used.}
\label{fig2_throughput vs Pb}
\end{figure}

Network throughput $R$ is plotted against $P_{b}$ in Fig. \ref{fig2_throughput vs Pb} to determine the best choice of number of UAVs ($n$) used in the considered square area of $20\times 20$ km$^{2}$. In addition, the impact of different numbers of $n$ on $R$ has been analyzed. The curves are generated based on random coordinates of UAVs with minimum distance among them so that the redundancy in the sensed data is minimized. Our approach optimally allocates transmission power $P_b$ among all UAVs. Initially $R$ increases significantly, because $R \propto P_i \,\hspace{2pt} \forall i\in \mathbb{N}_0$. However, beyond $n = 25$, there is no significant change in $R$. As a result, $n=25$ appears to be sufficient to provide the maximum coverage area. Although $R$ is more for higher values of $n$ but the expense of deploying such a huge number of UAVs is also higher.

\begin{figure}[h]
\begin{center}
\includegraphics[width=3.5in]{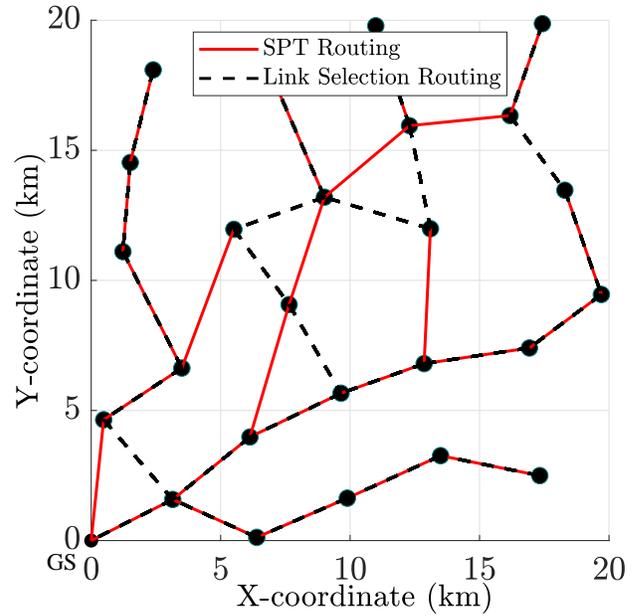}
\caption{Random UAV deployment for the desired square area with the best choice of number of UAVs discussed.}
\label{fig3_routing}
\end{center}
\end{figure}

Fig. \ref{fig3_routing} depicts the random placement of UAV to maximize the coverage area. The result from the solution of \text{($\bm{\mathbb{P}}1.1$)} is represented by red solid paths. On the other hand, black dashed paths represent the result of optimization subproblem \text{($\bm{\mathbb{P}}1.4$)}, which is the relaxed version of problem \text{($\bm{\mathbb{P}}1.2$)}.
\begin{figure}[h]
\centering
\includegraphics[width=3.8in]{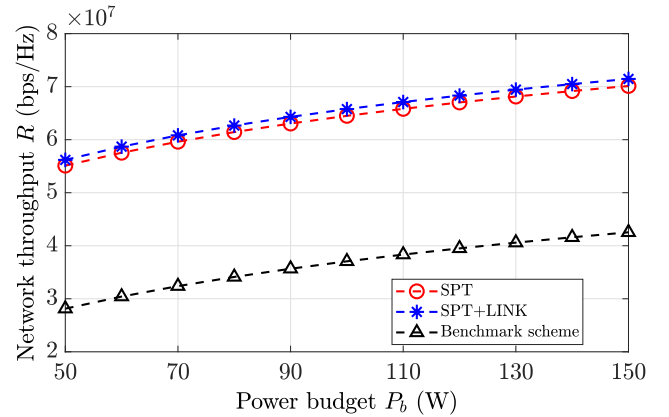}
\caption{Effect of $P_{b}$ on $R$ for different routing schemes compared with the benchmark scheme \cite{Chen_Multihop}.}
\label{fig4_throughput vs Pb_for_n_25}
\end{figure}

In Fig. \ref{fig4_throughput vs Pb_for_n_25}, the throughput $R$ achieved by both the subproblems are plotted with respect to $P_{b}$. It can be observed that the problem \text{($\bm{\mathbb{P}}1.4$)} significantly outperforms \text{($\bm{\mathbb{P}}1.1$)} because solution of power allocation subproblem is used in link selection subproblem, which further improves the throughput by interchanging the appropriate child-parent links. In this figure, we have also compared our work with the benchmark scheme investigated in \cite{Chen_Multihop}. It can be seen that the proposed schemes outperforms the benchmark scheme in terms of network throughput improvement. 

\section{Conclusion}
In this paper, we investigated the problem of maximizing the overall network throughput for UAV-assisted FANET via jointly optimizing multi-hop routing structure as well as power allocation. It is achieved by first computing the number of UAVs that can be deployed in a considered scenario. From the results, it can be found that $n = 25$ is the best choice for a given square area of $20\times 20$ km$^{2}$. The main formulated problem appears to be non-convex when optimizing both the variable jointly. In order to solve the problem efficiently it is decoupled into two subproblems. A global sub-optimal solution for allocating the power to each UAV is found from the power allocation subproblem. Then, the link selection subproblem is solved to further improve the network throughput using the solution of the first subproblem. Moreover, we have provided the simulation results to demonstrate the effectiveness of the proposed routing structure. Furthermore, we have also compared our work with a benchmark scheme, where the proposed schemes show a considerable performance improvement in maximizing the overall network throughput. Our proposed routing scheme can assist the UAV deployment in emergency search and rescue in disasters etc. Future extensions include optimal UAV deployment with different altitudes and the impacts of small-scale fading and interference among UAVs.

\appendices
\setcounter{equation}{0}
\setcounter{figure}{0}
\renewcommand{\theequation}{A.\arabic{equation}}
\renewcommand{\thefigure}{A.\arabic{figure}}

\section{Proof of Convexity}\label{AppA}
In this appendix, we provide the proof of convexity of $\sum_{i=1}^{n}R_{i} \triangleq R$

\begin{equation}\label{1stdiff}
\frac{\partial R_{i} }{\partial P_{i}} = \sum_{i=i}^{n}\frac{1}{\left ( 1+\frac{P_{i}h_{i}}{\sigma^{2}B} \right )} \frac{h_{i}}{\sigma^{2}},
\end{equation}

therefore, from (\ref{1stdiff}),
\begin{equation}\label{2nddiff}
\frac{\partial^2 R_{i} }{\partial P^2_{i}} = -\sum_{i=i}^{n}\frac{B}{\left(1+\frac{P_{i}h_{i}}{\sigma^{2}B} \right )^{2}} \left ( \frac{h_{i}}{\sigma^{2}B} \right )^2.
\end{equation}

Expression \eqref{2nddiff} is negative thus $R_{i}$ is a concave function of $P_{i}$ \cite{boyd2004convex}. Moreover, constraints $\sum_{i=1}^{n} P_{i}= P_{b}$ and $P_{i} \geq 0$ are linear over $P_{i}$, so they form convex set. Thus, \text{($\bm{\mathbb{P}}1.1$)} is a convex optimization problem \cite{boyd2004convex}.

\makeatletter
\renewenvironment{thebibliography}[1]{%
  \@xp\section\@xp*\@xp{\refname}%
  \normalfont\footnotesize\labelsep .5em\relax
  \renewcommand\theenumiv{\arabic{enumiv}}\let\p@enumiv\@empty
  \vspace*{-1pt}% NEW
  \list{\@biblabel{\theenumiv}}{\settowidth\labelwidth{\@biblabel{#1}}%
    \leftmargin\labelwidth \advance\leftmargin\labelsep
    \usecounter{enumiv}}%
  \sloppy \clubpenalty\@M \widowpenalty\clubpenalty
  \sfcode`\.=\@m
}{%
  \def\@noitemerr{\@latex@warning{Empty `thebibliography' environment}}%
  \endlist
}
\makeatother

\bibliographystyle{IEEEtran}
\bibliography{ref}
\end{document}